\documentstyle[12pt,aasms4]{article}
\begin{document}

\title{ The Effect of Weak Gravitational Lensing on the Cosmic Microwave
Background Anisotropy: Flat versus Open Universe Models}
\author{Enrique Mart\' \i nez-Gonz\'alez, Jose
L. Sanz}
\affil{Instituto de F\' \i sica de Cantabria, CSIC-Universidad de Cantabria,\\
Facultad de Ciencias, Avda. Los Castros s/n, 39005 Santander, Spain} 
\author{Laura Cay\'on\altaffilmark{1}}
\affil{Astronomy Department and Center for Particle Astrophysics,
University of California, Berkeley, CA 94720}  
\altaffiltext{1}{Permanent address: Instituto de F\'\i sica de Cantabria, 
CSIC-Universidad de Cantabria, Facultad de Ciencias, Avda. Los Castros s/n, 
39005 Santander, Spain}

\begin{abstract}
We have studied the effect of gravitational lensing on the Cosmic Microwave
Background (CMB) anisotropy in flat and open universes. We develop a 
formalism to 
calculate the changes on the radiation power spectrum induced by lensing in
the Newtonian and synchronous-comoving gauges. The previously
considered negligible  
contribution to the CMB radiation power spectrum of the 
anisotropic term of the lensing correlation is shown to be appreciable.
However, considering the nonlinear evolution of the matter power spectrum
produces only slight differences on the results based on linear evolution. 
The general conclusion for flat as well as open universes is
that lensing slightly smoothes the radiation power spectrum. 
For a given range of multipoles the effect of lensing increases 
with $\Omega$
but for the same acoustic peak it decreases with $\Omega$. The
maximum contribution of lensing to the radiation power spectrum for $l\leq
2000$ is
$\sim 5\%$ for $\Omega$ values in the range $0.1-1$.

\end{abstract}

\section{Introduction}\label{intr}

Cosmic Microwave Background (CMB) temperature anisotropies, detected for the
first time by Smoot et al. (1992) with the COBE-DMR experiment, are believed 
to be generated by the interaction of matter density perturbations and 
radiation to first order in perturbation theory. Numerical codes used to solve
the  linearized Einstein-Boltzmann coupled equations  are able to calculate the
radiation power spectrum with an accuracy better than $1\%$ (See e.g. 
Sugiyama 1996, Seljak \& Zaldarriaga 1996, Bond 1995). Nonlinear density 
perturbations make a small contribution through the Rees-Sciama effect which,
except for the case of reionization, can be constrained to be $\lesssim 1\%$
(Mart\'\i nez-Gonz\'alez, Sanz and Silk 1992, Sanz et al. 1996, Seljak 1996a, 
Tuluie, Laguna \& Anninos 1996). However, the 
effect of gravitational lensing on the CMB anisotropies, not included in the 
numerical codes, may appreciably affect the radiation power spectrum.

Many groups have studied the lensing of the microwave photons using different
analytical and numerical approaches
(Blanchard \& Schneider 1987; Cole \& Efstathiou 1989; Sasaki 1989; Tomita \&
Watanabe 1989; Linder 1990a, b; Cay\'on, Mart\'\i nez-Gonz\'alez \& Sanz 1993a,
b; Fukushige, Makino \& Ebisuzaki 1994; Seljak 1996b).
They arrive at
different conclusions about the importance of the effect: the result depends on
the particular cosmological model considered and on the asumptions made in the
calculation. Cay\'on et al. 1993a,b present the formalism to obtain the
lensing of the microwave photons by the large scale matter distribution in
a flat universe with null/non-null cosmological constant. However they
erroneously used the photon deflection angle instead of the photon angular
excursion on the last scattering surface relative to its observed value, which
leads to a factor of a few overestimate of the relative dispersion between two
photons (Seljak 1996b, Mu\~noz \& Portilla 1996).   
Some of the previous studies have used models that may not be
a realistic representation of the large-scale structure observed (e.g. the
models used in Fukushige et al. 1994). Another relevant ingredient of the
calculations is to appropriately account for the evolution of matter density
perturbations. Recently, Seljak (1996b) has done a relevant step in solving
those shortcomings of previous studies. Based on a power spectrum approach
he includes linear and nonlinear regimes of the matter evolution in realistic
cosmological models and generalizes the formalism to open universes. However,
results on the radiation power spectrum are not presented for open universes.
Moreover, the nonlinear power spectrum evolution considered in that paper is
not valid for spectral indexes $n<-1$ (as in the case of CDM for small scales)
and for $\Omega<1$ universes (Peacock and Dodds 1996).  

In this paper we present a formalism to calculate the lensing effect in flat
and open cosmological models and in two different gauges. Except for velocity
and acceleration terms associated to the observer and the source which
either do not contribute or the contribution is negligible, we show 
that the equations which provide the
lensing effect are the same for the conformal Newtonian and 
synchronous-comoving gauges.  
Results for the effect of lensing on the radiation power spectrum are
presented for CDM models with $0.1\leq\Omega\leq 1$. We consider linear and
nonlinear evolution for the matter power spectrum.
The structure of the paper is as follows: in section II
we describe the formalism to calculate the gravitational lensing effect. The
results obtained for CDM open models are presented in section III. Finally,
the main conclusions are related in section IV.  
 
\section{Formalism}\label{form}

\subsection{Geodesics in the conformal Newtonian gauge}
We will consider the propagation of photons from recombination to the present
time, the universe being a perturbed Friedmann model with a dust ($p = 0$)
matter content. We shall not consider a cosmological $\Lambda$-term, but the
generalization to include $\Lambda \neq 0$ is very easy. For scalar 
perturbations, the metric in the conformal Newtonian gauge is given in terms 
of a single potential $\phi (\tau, \vec{x})$ as follows
\begin{equation}
ds^2 = a^2(\tau )[-(1 + 2\phi )d{\tau }^2 + (1 - 2\phi ){\gamma }^{-2}
{\delta }_{ij}dx^idx^j], \ \ \ 
\gamma = 1 + \frac{k}{4}{|x|}^2, \ \ \  
\end{equation}
\noindent we take units such that $c = 8\pi G = a_o = 2H_o^{-1} = 1$ and 
$k/(4\mid 1-\Omega\mid) = 0, -1 , +1$ denote the flat, open and closed 
Friedmann background 
universe. The gravitational potential satisfies the Poisson equation
\begin{equation}
({\nabla }^2 + 3k)\phi = \frac{1}{2}{\rho}_ba^2\delta,
\end{equation}
\noindent where $\delta$ is the density perturbation.
The Green's function associated to the previous equation can be found in the
literature (D'Eath 1976, Traschen and Eardley 1986). We are interested in the
effect of gravitational lensing on high multipoles ($l\sim 10^3$) of the 
CMB. Only the smaller scales are contributing to such effect, so curvature 
will show related to the angular distance. In fact, the Green's function on 
such scales can be approximated by 
\begin{equation}
G(\vec{x}, {\vec{x}}^{'}) \simeq - \frac{1}{4\pi} 
{|\vec{x}-{\vec{x}^{'}}|}_\Omega^{-1},
\end{equation}
\noindent where the distance between the two points $\vec{x} = \lambda \vec{n},
{\vec{x}^{'}} = \lambda^{'}{\vec{n}^{'}}$ (being $\vec{n}$ and ${\vec{n}^{'}}$ 
two unit vectors in the  directions of observation) is given by the equation
($\lambda\approx\lambda^{'}$)
\begin{equation}
{|\vec{x} - {\vec{x}^{'}}|}_{\Omega}\simeq [s^2 + s^{'2} - 
   2ss^{'}\cos \alpha ]^{1/2},\ \ \  
s\equiv \frac{\lambda}{1 - (1 - \Omega ){\lambda}^2},\ \ \ 
\cos \alpha \equiv \vec{n} \cdot {\vec{n}^{'}}.
\end{equation}
On the other hand, after a straightforward calculation, the geodesic equation
associated to the metric (1) gives the following equation for the vector $s^i
\equiv \frac{k^i}{k^0} = \frac{d\,x^i}{d\,\tau}$
\begin{equation}
\frac{d\, s^i}{d\,\tau} = k{\gamma}^{-1} \Bigl[(\vec{x}\cdot \vec{s})s^i - 
\frac{1}{2}{\gamma}^2x^i\Bigr] - 2k\gamma \phi x^i + 2\Bigl[\frac{d\,\phi}
{d\,\tau} + 2\, (\vec{\nabla}\phi \cdot \vec{s})\Bigr]s^i - 2{\gamma}^2 
(\nabla \phi)^i.
\end{equation}
\noindent Assuming a perturbation scheme ("weak lensing"), this equation can 
be integrated in the form
\begin{equation}
\vec{x} = \lambda \vec{n} + \vec{\epsilon},
\end{equation}
\noindent where $\vec{n}$ is the direction of observation and $\lambda$ is the 
distance to the photon for the background metric, i.e.
\begin{equation}
\lambda = {\tau}_o - \tau \ \ \ (k = 0),\ \ \ 
\lambda = {(1 - \Omega )}^{-1}\tanh[(1 - \Omega )( {\tau}_o - \tau )]\ \ \ 
(k = -1) .
\end{equation}
The perturbation $\vec \epsilon$ can be decomposed in a term parallel to
$\vec{n}$ and a term orthogonal to such a direction ${\vec{\alpha }}_{\bot}$. 
The last term satisfies the following differential
equation when parametrized by $\lambda$
\begin{equation}
\frac{d^2{\vec{\alpha}}_{\bot}}{d\,{\lambda}^2} + \frac{k}{2\,\gamma}\Bigl[-
\lambda \frac{d\,{\vec{\alpha}}_{\bot}}{d\,\lambda } + 
{\vec{\alpha}}_{\bot}\Bigr] = - 2\,\vec{\nabla}_{\bot}{\phi} ,
\end{equation}
\noindent where $(\vec{\nabla}_{\bot}{\phi})^i \equiv ({\delta }^{ij} -
n^in^j){\partial\phi/\partial{n^j}}$. The solution to the previous equation 
with the initial conditions: ${\vec{\alpha}}_{\bot}(\lambda = 0) = 0 =
\frac{d\,{\vec{\alpha}}_{\bot}}{d\,\lambda }(\lambda = 0)$ is
\begin{equation}
\vec{\alpha}_{\bot} = - 2\int_0^{\lambda}d{\lambda}^{'}
W(\lambda , {\lambda}^{'})\vec{\nabla}_{\bot}\phi ({\lambda}^{'}, \vec{x} =
{\lambda}^{'}\vec{n})\ \ \
\end{equation}
\noindent where $W(\lambda, {\lambda}^{'})$ is a window function
\begin{equation}
a(\lambda) = \frac{(1 - \lambda )^2}{1 + k{\lambda}^2/4},\ \ \ 
W(\lambda, {\lambda}^{'}) = (\lambda - {\lambda}^{'})
\frac{1 + k\lambda {\lambda}^{'}/4}{1 + k{{\lambda}^{'}}^2/4}.
\end{equation}
For photons that are propagated from recombination, ${\lambda}_r = [1 - (1 +
\Omega z_r)^{-1/2}][1 - (1- \Omega)(1 + \Omega z_r)^{-1/2}]^{-1}$ with 
$z_r \simeq 10^3$, to the observer, ${\lambda}_o = 0$, the lensing vector 
$\vec{\beta}$ is defined in the usual way (see Figure 1)
\begin{equation} 
\vec{\beta} \equiv \vec{n} - \frac{\vec{x}_r - \vec{x}_o}
{|\vec{x}_r - \vec{x}_o|},
\end{equation}
\noindent so we find $\vec{\beta} = - \frac{1}{{\lambda}_r}
\vec{\alpha}_{\bot}({\lambda}_r)$ and the final result, taking into account
equations (9, 10), is
\begin{equation}
\vec{\beta} = - 2\int_0^1d{\lambda}
W(\lambda )\vec{\nabla}_{\bot}\phi ({\lambda}, \vec{x} =
{\lambda}\vec{n})\ \ \
W(\lambda ) = (1 - {\lambda})
\frac{1 - (1 - \Omega )\lambda }{1 - (1 - \Omega ){\lambda}^2},
\end{equation}
because ${\lambda}_r \simeq 1$ for $\Omega z_r\geq 10^2$.

The lensing vector for a flat universe has been
given by Kaiser (1992). For the open case, Pyne \& Birkinshaw (1996) and 
Seljak (1996b) have used a window  function $W$ that agrees after a
straightforward calculation with our equation (12).

\subsection{Geodesics in the conformal synchronous-comoving gauge}
Once we have obtained the expression for the trajectory of the photon in the
conformal Newtonian gauge, it is easy to calculate everything in the conformal
synchronous-comoving gauge. The infinitesimal transformation connecting both 
gauges is 
\begin{equation}
{\tau}^{'} = \tau + {\epsilon}^0(\tau, \vec{x}),\ \ \    
x^{i'} = x^{i} + {\epsilon}^i(\tau, \vec{x}), \ \ \ \ \ 
{\epsilon}^0 = \frac{2}{a^3{\rho}_b}\frac{\partial}{\partial \tau}(a\phi),\ \ \
{\epsilon}^i = \frac{2{\gamma}^2}{a^3{\rho}_b}\vec{\nabla} (a\phi).
\end{equation}
\noindent The expressions for ${\epsilon}^0$ and ${\epsilon}^i$ can be 
obtained taking
into account that in the synchronous-comoving gauge one has zero velocity
($v^{i'} = 0$), i. e.
\begin{equation}
u^0 = (1 - \dot{\epsilon}^0)u^{0'},\ \ \ u^i = - \dot{\epsilon}^iu^{0'},\ \ \ 
v^i = - \dot{\epsilon}^i, \ \ \ \dot{} \equiv \frac{\partial}{\partial \tau},
\end{equation}
\noindent and the metric has the following components: 
$g_{0'0'} = - a^2({\tau}'), g_{0'j'} = 0$. So, 
\begin{equation}
\dot{\epsilon}^0 + \frac{\dot{a}}{a}{\epsilon}^0 = \phi,\ \ \ 
{\epsilon}_{,j}^0 - {\gamma}^{-2}\dot{\epsilon}_j = 0
\end{equation}
\noindent and integrating the last equations we get the result mentioned 
above for $({\epsilon}^0, {\epsilon}^i)$.
\noindent Moreover, the metric in the conformal synchronous-comoving 
gauge reads
\begin{equation}
ds^2 = a^2({\tau}^{'})\Bigl\{-d{\tau}^{'2} + {\gamma}^{-2}\Bigl[(1 - 2\phi
-2\frac{\dot{a}}{a}{\epsilon}^0 + k{\gamma}^{-1}\vec{x}^{'}\cdot
\vec{\epsilon}){\delta}_{ij} - {\epsilon}_{i,j} - {\epsilon}_{j,i}\Bigr]
dx^{i'}dx^{j'}\Bigr\}.
\end{equation}
                                                                              
\noindent This last expression for $k = 0$ agrees with the one given by 
Sachs \& Wolfe (1967). By changing the gauge, the new lensing vector 
$\vec{\beta}^{'}$ is given by an equation similar to (11), so we obtain
\begin{equation}
\vec{\beta}^{'} = \vec{\beta} - \frac{1}{{\lambda}_r}(\vec{\epsilon}_{\bot r} -
\vec{\epsilon}_{\bot o}) - \Bigl(\frac{d\vec{\epsilon}_{\bot}}
{d\tau}\Bigr)_o,\ \ \ 
\end{equation}
\noindent where ${\epsilon}^0, \vec{\epsilon}$ are given by equation (13). 
The velocity of the fluid in the conformal Newtonian 
gauge is given by $\vec{v} = - \frac{\partial \vec{\epsilon}}{\partial\tau}$,
from which $\vec\epsilon=-(a/\dot{a} f)\vec v$ ($f\equiv d\ln D/d\ln a$,
$D(a)$ being the growing mode).
Taking this into account one can easily understand that the new terms 
appearing in equation (17)
can be interpreted as Doppler contributions at recombination and at the 
observer and an acceleration term at the observer. For a flat model 
($k = 0$), we explicitly have
\begin{equation}
\vec{\beta}^{'} \simeq \vec{\beta} - \frac{1}{2}{\biggl[\frac{\vec{v}_r -
\vec{v}_o}{(1 + z_r)^{1/2}} + \vec{a}_o\biggr]}_{\bot},
\end{equation}
\noindent where the linear gravitational potential, $\phi
(\vec{x})$, is time-independent and $\vec{v}_o = -
\frac{1}{3}\vec{\nabla}{\phi}_o$, $\vec{v}_r = - \frac{1}{3}(1 +
z_r)^{-1}\vec{\nabla}{\phi}_r$ and $\vec{a}_o = (\frac{d\vec{v}}{d\tau})_o$. 
The ratio of these terms, as they appear in equation (18), to the angular
scale is negligible ($\vec v_o$ is given by the Doppler velocity respect to
the CMB and $\vec a_o$ can be estimated from our local infall towards either
the Virgo cluster or the Great Attractor). A similar reasoning can be applied 
to open universes.
Therefore, the lensing vector $\vec{\beta}$ in the synchronous-comoving gauge
(that is the appropriate one from the point of view of the observations) is
approximately given by $\vec{\beta}$, as defined by equation (12).

\subsection{The influence of weak gravitational lensing on the ${C_l}^{'}$s}
The correlation function $\bar{C}(\theta)$ including gravitational lensing is
calculated as the average
\begin{equation}
\bar{C}(\theta) = \langle \Delta (\vec{n} + \vec{\beta}(\vec{n}))\,
\Delta (\vec{n}^{'} + \vec{\beta}(\vec{n}^{'}))\rangle,
\end{equation}
\noindent where $\Delta (\vec{n})$ is the temperature anisotropy field, 
$\vec{n}$ and $\vec{n}^{'}$ are two directions such that
$\vec{n} \cdot \vec{n}^{'} = \cos \theta $. By introducing 2D-Fourier
components of the temperature anisotropies ${\Delta}_{\vec{q}}$ and assuming
that the anisotropies, $\Delta$, and the lensing vector, $\vec{\beta}$ are 
uncorrelated, we obtain
\begin{equation}
\bar{C}(\theta) = \frac{1}{2\pi}\int_0^{\infty}dq\,qP_{\Delta}(q)
\langle J_0(q\nu)\rangle,
\end{equation}
\noindent where $J_0$ is the Bessel function, 
$\nu \equiv |\vec{n} - \vec{n}^{'} + \vec{\beta}(\vec{n}) - 
\vec{\beta}(\vec{n}^{'})|$ 
and $P_{\Delta}$ is the 2D-power spectrum of the radiation field: 
$\langle \Delta _{\vec{q}}\Delta _{\vec{q}^{'}}^{*}\rangle = 
P_{\Delta}(q){\delta}^2(\vec{q} - \vec{q}^{'})$. 
On the other hand, assuming
weak gravitational lensing, i.e. on the average the relative lensing vector is 
very small as compared to the angle $\theta$, we can make a series expansion in
the previous equation obtaining
\begin{equation}
\bar{C}(\theta) - C(\theta) = \frac{1}{2{\theta}^2}\biggl\{\Bigl[Q_k^k - 
Q_{ij}\frac{{\theta}^i{\theta}^j}{{\theta}^2}\Bigr]\theta \frac{dC(\theta)}
{d\theta}
+ \Bigl[Q_{ij}\frac{{\theta}^i{\theta}^j}{{\theta}^2}\Bigr]{\theta}^2
\frac{d^2C(\theta)}{d{\theta}^2}\biggr\}.
\end{equation}
\noindent $Q_{ij}$ is the bending correlation matrix
\begin{equation}
Q_{ij} \equiv \langle [{\beta}_i(\vec{n}) - {\beta}_i(\vec{n}^{'})]
                      [{\beta}_j(\vec{n}) - {\beta}_j(\vec{n}^{'})]\rangle,
\end{equation}
and can be decomposed into the trace and an anisotropic component
\begin{equation}
Q_k^k \equiv 2{\sigma}^2(\theta),\ \ \  
\xi (\theta) \equiv Q_{ij}\frac{{\theta}^i{\theta}^j}{{\theta}^2} - 
{\sigma}^2,
\end{equation} 
\noindent where $\sigma (\theta)$ is the bending dispersion and $\xi(\theta)$ 
is the 
anisotropic correlation ($\xi(\theta)$ corresponds to C$_{gl,2}(\theta)$ in
Seljak 1996b). Therefore equation (21) can be rewritten as
\begin{equation}
\bar{C}(\theta ) - C(\theta ) =
\frac{{\sigma}^2}{2}\Bigl[\frac{d^2C(\theta)}{d{\theta}^2} + \frac{1}{{\theta}}
\frac{d\,C(\theta)}{d{\theta}}\Bigr] +
\frac{{\xi}}{2}\Bigl[\frac{d^2C(\theta)}{d{\theta}^2} - \frac{1}{{\theta}}
\frac{d\,C(\theta)}{d{\theta}}\Bigr]. 
\end{equation} 

On the other hand, taking into account the expansion
\begin{equation}
C(\theta ) = \frac{1}{4\pi }{\sum}_l (2l + 1)C_lP_l(\cos \theta )
\end{equation}
\noindent and the approximation $P_l(\cos \theta ) \simeq J_0(l\theta )$ for 
$l\gg 1$, we get
\begin{equation}
C_l \simeq 2\pi\int_0^2\,d\theta \,\theta \,C(\theta )J_0(l\theta )\ \ \ 
(l\gg 1).
\end{equation}
\noindent From equations (24) and (26)
\begin{equation}
\bar{C}_l - C_l = - \frac{1}{4}{\sum }_{l^{'}}(2l^{'} + 1)l^{'2}C_{l^{'}}
\int_0^2\,d\theta \,\theta\, J_0(l\theta )[{\sigma}^2{(\theta)
J_0(l^{'}\theta ) - {\xi}}(\theta)J_2(l^{'}\theta )], \ \ \  (l\gg 1).
\end{equation}

The next step is the calculation of the dispersion and correlation of the
lensing vector as a function of the power spectrum $P(a, k)$ defined by
\begin{equation}
\langle {\delta}_{\vec k}(a){\delta}^*_{\vec k^{'}}(a)\rangle \equiv P(a, k)
{\delta}^3(\vec{k} - \vec{k}^{'}).
\end{equation}
From equation (12) one can obtain
\begin{equation}
\langle {\beta}_i{\beta}_j\rangle = 4D_iD_j\int_0^1d\lambda
\frac{W(\lambda )}{\lambda}\int_0^1d{\lambda}^{'} \frac{W({\lambda}^{'})}
{{\lambda}^{'}}C_{\phi}(\lambda ,{\lambda}^{'}, r),
\end{equation}
\noindent where $D_i \equiv ({\delta}^k_i - n^kn_i){\partial/\partial{n^k}}$ 
and $C_{\phi}(\lambda ,{\lambda}^{'}, r)$ is the correlation of the
gravitational potential at two different times. If one assumes Limber's
approximation (see also Kaiser 1992), i.e. only a small region $r$ 
with ${\lambda}^{'} \simeq \lambda$ is contributing, the previous equation 
can be approximated by
\begin{equation}
\langle {\beta}_i{\beta}_j\rangle = 8D_iD_j\int_0^1d\lambda
\biggl[\frac{W(\lambda )}{\lambda}\biggr]^2\int_{\theta s}^{\infty} dr
r{(r^2 - {\theta}^2 s^2)}^{-1/2}[1 - (1 - \Omega ){\lambda}^2]
C_{\phi}(\lambda , r).
\end{equation}
\noindent Notice that the correlation depends only on a single time 
and $s$ is given by equation (4). Introducing the 
power spectrum $P_{\phi}(a, k)$, the last expresion becomes
\begin{equation}
\langle {\beta}_i{\beta}_j\rangle = \frac{2}{\pi}D_iD_j\int_0^1d\lambda
\biggl[\frac{W(\lambda )}{\lambda}\biggr]^2[1 - (1 - \Omega
){\lambda}^2]\int_0^{\infty}\,dk\,k^{-1}P_{\phi}(\lambda, k)J_0(ks\theta ).
\end{equation}
$P_{\phi}(a, k)$ is given by the Poisson equation (2) for scales
($k^2 \gg 12(1 -\Omega )$)
\begin{equation}
P_{\phi}(a, k) \simeq {(\frac{6\Omega }{a})}^2k^{-4}P(a, k),
\end{equation}
\noindent where $P(a, k)$ is the power spectrum associated to the matter
perturbations. In the linear regime: $P(a, k) = D(a)P(k)$, $D(a)$ being the
growing mode  normalized to the present time (see Peebles 1980).

Finally, calculating the derivatives that appear in equation
(31) and applying equations (22, 23) one can obtain
\begin{equation}
{\sigma}^2(\theta) =
\frac{72{\Omega}^2}{\pi}\int_0^{\infty}\frac{dk}{k}\int_0^1d\lambda\,
{\biggl[\frac{W(\lambda )}{a}\biggr]}^2
\frac{P(a, k)}{1 - (1 - \Omega){\lambda}^2}
\Bigl[1 - J_0 + \frac{1}{2}\sin^2\theta J_0 - {\sin}^2\frac{\theta}{2}J_2
\Bigr] ,
\ \ \ 
\end{equation}
\begin{equation}
\xi (\theta) = {\sigma}^2 + 
\frac{36{\Omega}^2}{\pi}\int_0^{\infty}\frac{dk}{k}\int_0^1d\lambda\,
{\biggl[\frac{W(\lambda )}{a}\biggr]}^2
\frac{P(a, k)}{1 - (1 - \Omega){\lambda}^2}
\Bigl[(\cos \theta - 3)(1 - J_0 - J_2) - {\sin}^2\theta J_0\Bigr] ,\ \ \
\end{equation}
\noindent where the argument of the Bessel functions $J_0$ and
$J_2$ is $ks\theta$.
Notice that the behaviour of
$\sigma(\theta)$ and $\xi^{1/2}(\theta)$ for small $\theta$ is linear:
\begin{equation}
{\sigma(\theta)\over \theta}\rightarrow a\theta \ ,\ \ \
{\xi^{1/2}(\theta)\over \theta}\rightarrow b\theta \ ,\ \ \
b\simeq \frac {a}{\sqrt{2}} \ ,
\end{equation} 
as will be shown by the numerical calculations presented in the next section.

\section{Results}\label{res}
With the formalism presented in the previous section, we have calculated the
dispersion of lensing $\sigma(\theta)$ and the anisotropic term of the 
correlation of
lensing $\xi^{1/2}(\theta)$ as given by equations (33-34). We assume a CDM 
model
with a primordial Harrison-Zeldovich spectrum, a Hubble parameter $h=0.5$ 
($H=100 h$km s$^{-1}$ Mpc$^{-1}$) and flat and open universe models. 
The radiation power spectrum not including lensing is normalized to the 2-year
COBE-DMR map as given by the analysis of Cay\'on et al. (1996) (this
normalization does not appreciably change with the 4-year data). 
However,   
since the lensing effect is generated by small scales, $<< 100$ Mpc, it might
be more sensible to use the normalization $\sigma_{8}=0.6,1,1.4$ for universes
with $\Omega=1,0.3,0.1$ following Viana and Liddle (1996). This normalization
is based on the cluster abundance (see also White, Efstathiou and Frenk 1993,
Eke, Cole and Frenk 1996). For the nonlinear evolution of the power spectrum
we use the recently improved fitting formula given by Peacock and Dodds
(1996). That formula is based on the Hamilton et al. (1991) scaling procedure
to describe the transition between linear and nonlinear regimes. It
accounts for the correction introduced by Jain, Mo and White (1995) for
spectra with $n\lesssim -1$ and applies to flat as well as to open universes.  

In figure 2 it is shown the relative dispersion 
$\sigma(\theta)/\theta$ and the anisotropic term $\xi^{1/2}(\theta)/\theta$ 
for three
values of the density parameter $\Omega=1, 0.3, 0.1$. Linear and nonlinear
matter evolutions have been considered for comparison. Discrepances between
the two regimes can be noticed at scales $\theta\lesssim 3'$. At scales
$\theta\gtrsim 6''$ $\sigma(\theta)/\theta$ as well as $\xi^{1/2}(\theta)/
\theta$ are below $20\%$ being slightly larger as $\Omega$ increases. Also, 
notice that $(\xi^{1/2}(\theta)/\theta)_{\theta\rightarrow 0}\simeq 
{1\over \sqrt 2}(\sigma(\theta)/\theta)_{\theta\rightarrow 0}$ in all cases 
(as expected from 
the considerations made in the previous section). Therefore, the anisotropic 
term 
should in principle be considered when calculating the distortions on the 
radiation power spectrum contrary to the isotropic approximation often made 
in the 
literature (we confirm this statement below). Lensing becomes negligible at
angular scales above a few degrees. 

The radiation power spectrum including and not including lensing is given in
figure 3a. It is clear from this picture that the effect of lensing is to
slightly smooth the main features appearing in the spectrum, in particular the
secondary acoustic peaks (also called Doppler or Sakharov). 
The relative
changes of the spectrum due to lensing as a function of the multipole $l$ are
shown in figures 3b,c. The changes produced grow with $l$ and in the same range
of $l$
increase with $\Omega$. Note, however, that for the same acoustic peak the
variation increases when $\Omega$ decreases and the reason for this is the
smaller scales involved for which the lensing effect is more effective. 
Considering nonlinear evolution does not change the lensing contribution to
the $C_l$ for $l\lesssim 2000$ as can be seen in figure 3b. 
We have also computed the contribution of the isotropic term to the $C_l$
coefficients and the result is shown in figure 3c. This contribution is
slightly smaller than the total effect and for some multipoles the discrepancy 
can be significant. Therefore, in general both terms, isotropic and
anisotropic, should be considered in the calculation of the radiation power
spectrum with lensing.

The effect of lensing can be as much as $\approx 2\%$ for multipoles
$l\lesssim 1000$
and $\approx 5\%$ for $l\lesssim 2000$. Therefore, if one wants to 
compute the radiation power spectrum for a particular cosmological model with
an accuracy better than $1\%$ such effect should be considered.
Bending of the microwave photons due to the large-scale structure should be
considered when analysing data provided by future very sensitive 
CMB experiments (e.g. COBRAS/SAMBA).

\section{Conclusions}\label{con}

A formalism has been developed to calculate the lensing effect on the primary
CMB radiation power spectrum. This formalism provides an expression for the 
lensing vector in flat and open
universes which is approximately  the same in both the conformal 
Newtonian and comoving-synchronous gauges. In particular, we give the window
function $W$ for open models in terms of the distance to the photon from the
observer.

The influence of gravitational lensing on the ${C_l}^{'}$s has been obtained
in terms of the bending dispersion $\sigma$ and anisotropic bending
correlation $\xi$. It is found that the contribution of $\xi$ to the lensing 
distortion of the
radiation power spectrum is smaller than that of $\sigma$. However this
contribution is not negligible and should be considered in the calculation of
the radiation power spectrum with lensing.

We use the recently improved fitting formula for the evolution of the
nonlinear matter power spectrum which provides an accuracy better than $12\%$ 
for the scales considered (Peacock and Dodds 1996). This improvement over
previous works (Peacock and Dodds 1993; Jain, Mo and White 1995) generates a
larger lensing dispersion at small scales for open models, as compared with 
Seljak (1996). In spite of this, the contribution of nonlinear evolution 
to the distortion of the radiation power spectrum is negligible. 

For flat as well as open universes, the effect of lensing is to slightly
smooth the primary radiation power spectrum of the CMB.
For a given range of multipoles the relative change of $C_l$ due to lensing 
increases 
with $\Omega$. However, for the same acoustic peak it decreases with $\Omega$.
The maximum contribution of lensing to the radiation power spectrum for 
$l\leq 2000$ is
$\approx 5\%$ for $\Omega$ values in the range $0.1-1$. Therefore,
the effect of lensing should be considered in analyses of CMB
anisotropy data provided by future very sensitive experiments.

\acknowledgements 
We would like to thank N. Sugiyama for providing us with the radiation power
spectrum for the models used in this paper.
EMG and JLS acknowledge financial support 
from the Spanish DGES, project PB95-1135-C02-02. LC has been supported in
part by a grant from NASA. EMG and JLS 
also acknowledge finantial support from the PECO contract of the EU ERBCIPDCT 
940019.

\newpage
\figcaption{Diagram describing the geometry and the angles involved in lensing
calculations on the CMB. Note that $\vec e\equiv (\vec x_r-\vec x_o)/|\vec
x_r-\vec x_o|$ as appears in equation (11).}

\figcaption{Ratios of the bending dispersion (a) and the anisotropic 
correlation (b)
to the angular distance, $\sigma(\theta)/\theta$, $\xi^{1/2}(\theta)/\theta$, 
as a function of the angular distance $\theta$ in units of arcmin. Solid, 
dashed and dotted lines correspond to $\Omega=1, 0.3, 0.1$ respectively.
Thick curves represent the results considering nonlinear evolution whereas 
thin ones outline the results from linear evolution.}

\figcaption{(a) Radiation power spectrum including (solid) and not 
including (dashed) lensing for $\Omega=1, 0.3, 0.1$. For $l=1000$ the lower
curve represents to $\Omega=1$ and the upper one to $\Omega=0.1$.
Relative change in the radiation power spectrum due to
lensing for $\Omega=1$ (solid), $\Omega=0.3$ (dashed) and $\Omega=0.1$
(dotted). (b) Thick curves represent the total effect (including nonlinear
evolution and the anisotropic term) whereas thin ones outline the result of
not considering nonlinear evolution.
(c) Thick curves represent the total effect whereas the thin ones
outline the result of not considering the anisotropic term.}
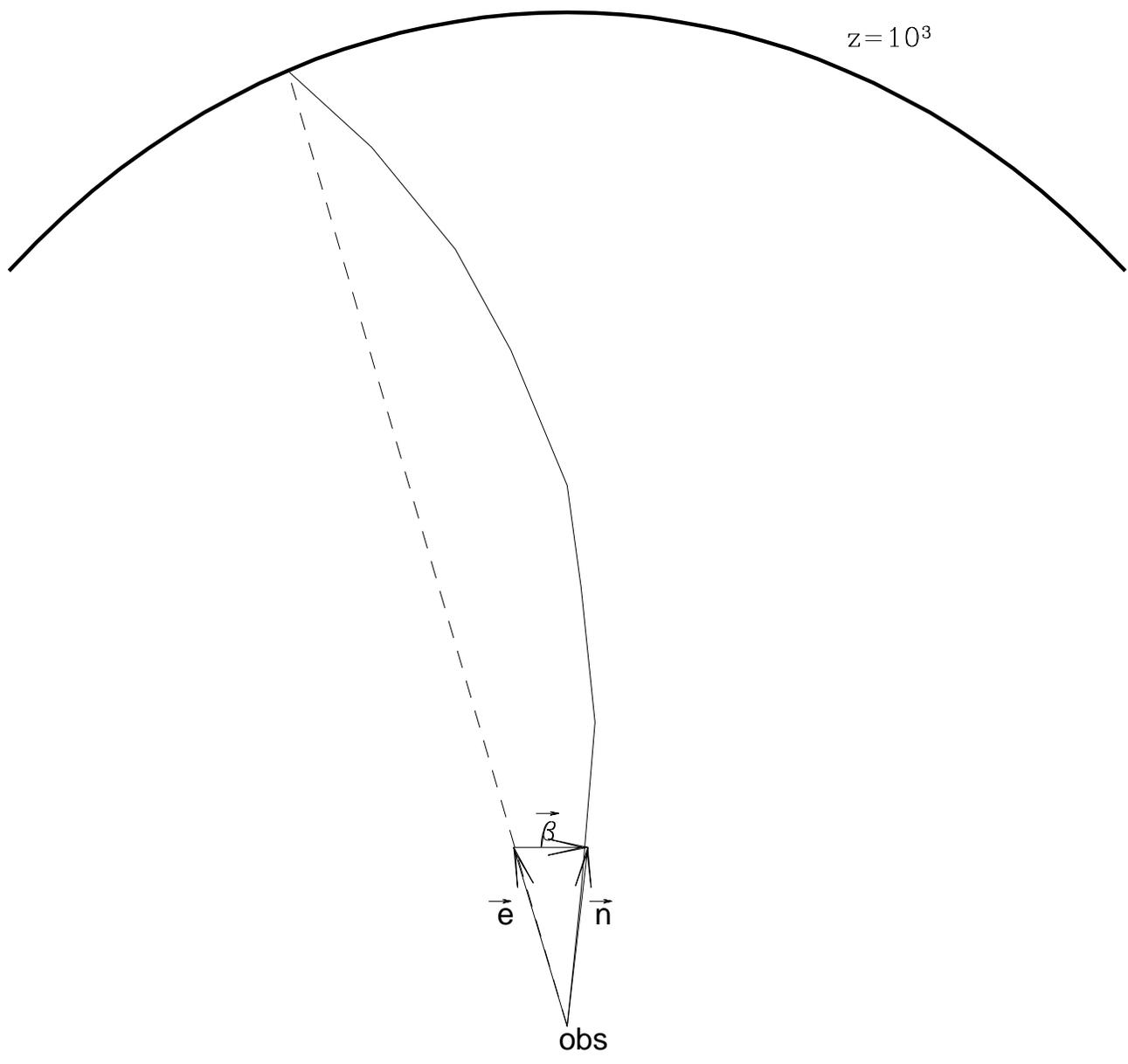
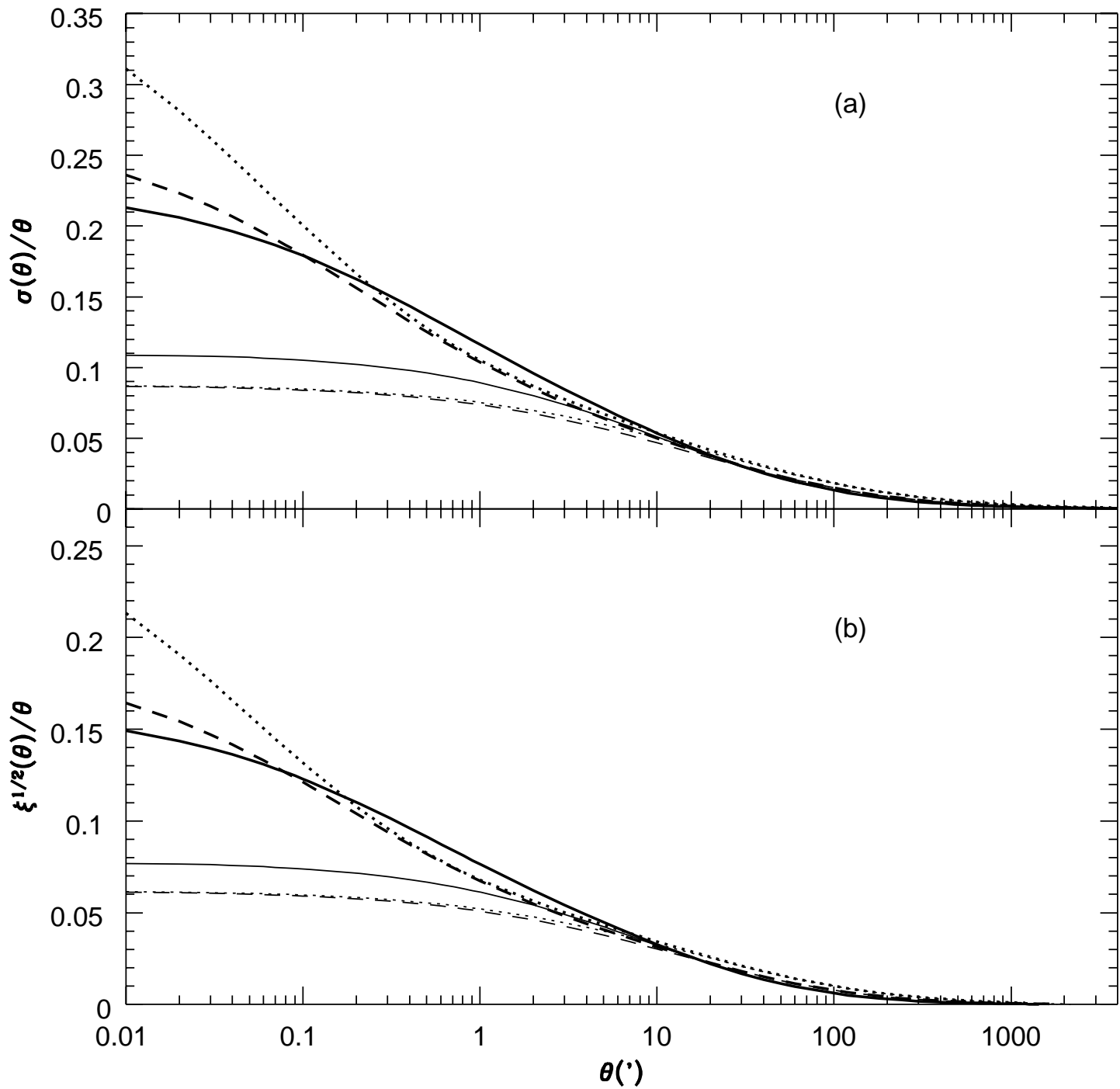
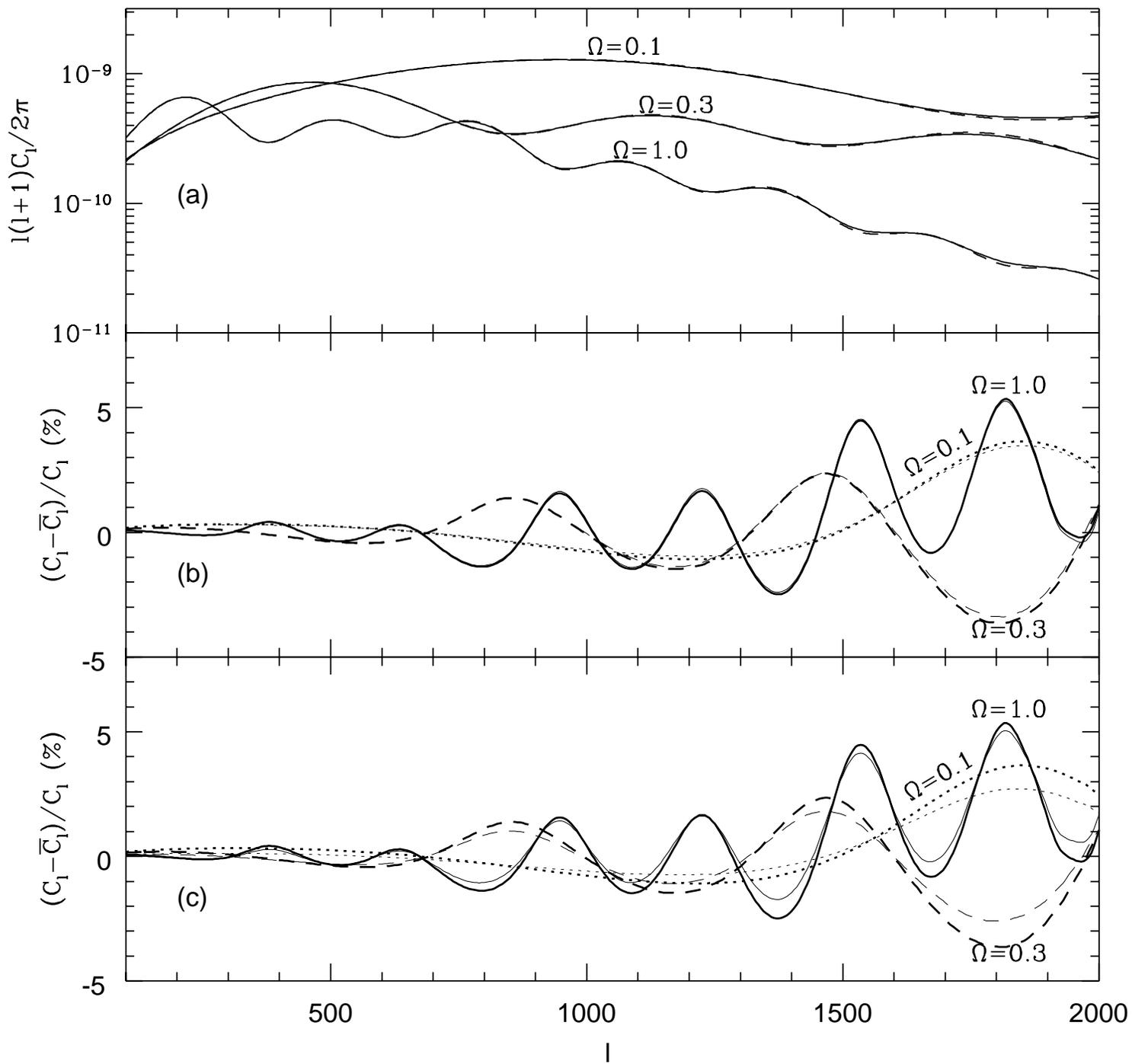
\end{document}